\documentclass[a4paper,fleqn,usenatbib,dvipdfmx]{mnras}

\usepackage{newtxtext,newtxmath}

\usepackage[T1]{fontenc}
\usepackage{ae,aecompl}

\usepackage{graphicx}	% Including figure files
\usepackage{amsmath}	% Advanced maths commands
\usepackage{xspace}
\usepackage{booktabs,caption}
\usepackage{threeparttable}
\usepackage{orcidlink}

\newcommand{\fdir}{./}
\newcommand{\msun}{M_\odot}
\newcommand{\zsun}{Z_\odot}
\newcommand{\rsun}{R_\odot}

\newcommand{\cago}{^{12}{\rm C}(\alpha,\gamma)^{16}{\rm O}}
\newcommand{\tsig}{3\sigma}

\title[Contribution of Pop III BH]{Contribution of Population III Stars to Merging Binary Black Holes}

\author[Ataru Tanikawa]{Ataru
  Tanikawa\orcidlink{0000-0002-8461-5517}$^{1,2}$\thanks{E-mail:
    tanik@g.fpu.ac.jp}, \\
  $^{1}$Center for Information Science, Fukui Prefectural University,
  4-1-1 Matsuoka Kenjojima, Eiheiji-cho, Fukui 910-1195, Japan\\
  $^{2}$Department of Earth Science and Astronomy, College of Arts and
  Sciences, The University of Tokyo, 3-8-1 Komaba, Meguro-ku, Tokyo
  153-8902, Japan\\}

\date{Accepted XXX. Received YYY; in original form ZZZ}

\pubyear{2017}

\begin{document}
\label{firstpage}
\pagerange{\pageref{firstpage}--\pageref{lastpage}}
\maketitle

\begin{abstract}

  A large number of mergers of binary black holes (BHs) have been
  discovered by gravitational wave observations since the first
  detection of gravitational waves 2015. Binary BH mergers are the
  loudest events in the universe, however their origin(s) have been
  under debate. There have been many suggestions for merging binary
  BHs. Isolated binary stars are one of the most promising origins. We
  have investigated the evolution of isolated binary stars ranging
  from zero metallicity (Population III stars or Pop III stars) to the
  solar metallicity by means of so-called rapid binary population
  synthesis simulation. We have found that binary BHs formed from
  isolated binary stars reproduce the redshift evolution of the merger
  rate density and the distribution of primary BH masses and mass
  ratios inferred by Gravitational-Wave Transient Catalog 3
  (GWTC-3). Pop III stars have a crucial role in forming merging
  binary BHs in so-called the pair instability mass gap. Note that we
  choose the conventional prescription of pair instability mass loss,
  based on the standard $\cago$ reaction rate. Finally, we have shown
  the redshift evolution of the rate density of pair instability
  supernovae, and have predicted that a few pair instability
  supernovae would be discovered in the next few years. The
  discoveries would validate our results of merging binary BHs.

  \if 0
An article usually includes an abstract, a concise summary of the work
covered at length in the main body of the article. 
\begin{description}
\item[Usage]
Secondary publications and information retrieval purposes.
\item[Structure]
You may use the \texttt{description} environment to structure your abstract;
use the optional argument of the \verb+\item+ command to give the category of each item. 
\end{description}
\fi

\end{abstract}

%\keywords{Suggested keywords}%Use showkeys class option if keyword
                              %display desired
\maketitle

%\tableofcontents

%~30 pages

\section{Introduction}

The origin of merging binary black holes (BHs) is one of the biggest
problems to solve in astronomy and astrophysics. In this paper, we
suggest that isolated binary stars can be their origin when we
consider metal-free or Population (Pop) III stars. In this section, we
introduce three ingredients in this paper. In section
\ref{sec:MergingBinaryBH}, we overview gravitational wave observations
of merging binary black BHs, and focus on binary BHs with $\sim
100\;\msun$ in the so-called pair instability (PI) mass gap. In
section \ref{sec:PairInstability}, we describe pair instability
supernovae (PISNe) forming PI mass gap. In section \ref{sec:PopIII},
we introduce Pop III stars, and briefly describe how Pop III stars
produce binary BHs in the PI mass gap. Finally, we shortly summarize
this section.

\subsection{Merging binary black holes}
\label{sec:MergingBinaryBH}

Gravitational waves are ripples in spacetime generated by accelerated
masses, and natural consequences of general relativity. Long after the
presence of gravitational waves was predicted 1916 by Albert Einstein,
it was indirectly confirmed by the orbital decay of the Hulse-Taylor
binary pulsar \citep{1975ApJ...195L..51H}. About 100 years after the
Einstein's prediction, gravitational waves have been directly detected
by Laser Interferometer Gravitational-wave Observatory (LIGO), and the
gravitational waves are generated by merging binary BHs
\citep{2016PhRvL.116f1102A}. This is also the first discovery of binary
BHs and their mergers. Gravitational wave signals have been detected
about 90 times until 2021 \citep{2019PhRvX...9c1040A,
  2021PhRvX..11b1053A, 2023PhRvX..13a1048A}, and most of them have
been generated by merging binary BHs except for a few events
\citep{2017PhRvL.119p1101A, 2020ApJ...892L...3A,
  2021ApJ...915L...5A}. This means that merging binary BHs are the
loudest objects in the LIGO band at a gravitational wave frequency of
$10$-$100$ Hz.

The discoveries of merging binary BHs pose a question of the origin of
such binary BHs. There are many suggestions as follows. Isolated Pop I
(or solar-metal) stars and Pop II (or subsolar-metal) stars produce
merging binary BHs through common envelope evolution
\citep{1998ApJ...506..780B, 2016Natur.534..512B, 2017NatCo...814906S,
  2017PASA...34...58E, 2017MNRAS.472.2422M, 2018MNRAS.481.1908K},
chemically homogeneous evolution \citep{2016MNRAS.458.2634M,
  2016A&A...588A..50M}, and stable mass transfer
\citep{2017MNRAS.471.4256V, 2019MNRAS.490.3740N,
  2021A&A...647A.153B}. Isolated Pop III (zero-metal) stars can
contribute to merging binary BH populations
\citep{2004ApJ...608L..45B, 2014MNRAS.442.2963K, 2016MNRAS.460L..74H,
  2017MNRAS.468.5020I, 2017MNRAS.471.4702B, 2020MNRAS.498.3946K,
  2021ApJ...910...30T, 2021MNRAS.505.2170T, 2021MNRAS.501L..49K,
  2021PTEP.2021b1E01K, 2021MNRAS.504L..28K, 2022PASJ...74..521T,
  2023MNRAS.524..307S, 2023arXiv230315511C}. Binary BHs can be formed
through dynamical capture in dense stellar clusters, such as globular
clusters \citep{2000ApJ...528L..17P, 2006ApJ...637..937O,
  2008ApJ...676.1162S, 2010MNRAS.407.1946D, 2011MNRAS.416..133D,
  2013MNRAS.435.1358T, 2014MNRAS.440.2714B, 2016PhRvD..93h4029R,
  2016ApJ...824L...8R, 2017PASJ...69...94F, 2017MNRAS.464L..36A,
  2017MNRAS.469.4665P, 2018MNRAS.480.5645H, 2018PhRvL.120o1101R,
  2018ApJ...855..124S, 2019PhRvD.100d3027R, 2020PASA...37...44A,
  2020ApJ...903...45K, 2021MNRAS.504.5778W, 2022MNRAS.509.4713W,
  2021ApJ...907L..25W}, open star clusters \citep{2010MNRAS.402..371B,
  2014MNRAS.441.3703Z, 2016MNRAS.459.3432M, 2017MNRAS.467..524B,
  2018MNRAS.473..909B, 2018MNRAS.481.5123B, 2019MNRAS.483.1233R,
  2019MNRAS.486.3942K, 2019MNRAS.487.2947D, 2020ApJ...902L..26F,
  2020MNRAS.495.4268K, 2020MNRAS.497.1043D, 2020MNRAS.498..495D,
  2020ApJ...898..152S, 2020ApJ...902L..26F, 2021MNRAS.500.3002B,
  2021MNRAS.503.3371B, 2021MNRAS.507.3612R, 2021MNRAS.502.4877S,
  2021ApJ...913L..29F, 2021ApJ...920..128A, 2022MNRAS.512..884R,
  2022MNRAS.511.4060K, 2022MNRAS.516.3266K}, Pop III clusters
\citep{2020MNRAS.495.2475L, 2020ApJ...903L..40L, 2021MNRAS.501..643L,
  2021MNRAS.506.5451L, 2022MNRAS.515.5106W}, and galactic centers
\citep{2009MNRAS.395.2127O, 2016ApJ...831..187A, 2017MNRAS.464..946S,
  2018MNRAS.474.5672L, 2018ApJ...866...66M, 2019ApJ...881...20R,
  2019ApJ...885..135T, 2019ApJ...876..122Y, 2020ApJ...891...47A,
  2020ApJ...898...25T, 2020ApJ...891...47A, 2021MNRAS.505..339M,
  2021ApJ...907L..20T, 2021ApJ...908..194T,
  2021Symm...13.1678M}. Hierarchical triple/quadruple stellar systems
can accelerate binary BH mergers \citep{2014ApJ...781...45A,
  2017ApJ...836...39S, 2018ApJ...863....7R, 2019MNRAS.486.4781F,
  2020ApJ...898...99H, 2022MNRAS.511.1362T,
  2023arXiv231105393L}. These BHs can be primordial BHs
\citep{2016PhRvL.117f1101S, 2021arXiv210503349F}. The BH may come not
from a single origin but from multiple origins
\citep{2021MNRAS.507.5224B, 2021JCAP...03..068H, 2021ApJ...913L...5N,
  2021ApJ...910..152Z}.

A large number of merging binary BHs enables us to assess the origin
of merging binary BHs statistically, using BH masses, BH spins,
positions, and redshifts. In particular, the mass distribution primary
(or heavier) BHs is used most frequently, because primary BH masses
are determined with an accuracy of $\sim 10$ \%
\citep{2019PhRvX...9c1040A, 2021PhRvX..11b1053A,
  2023PhRvX..13a1048A}. Note that effective spins of merging binary
BHs (or combinations of two BHs' spin components aligned to binary
BHs' spin angular momenta) contain much larger relative errors than
primary BH masses, since their typical values are zero
\citep{2019PhRvX...9c1040A, 2021PhRvX..11b1053A,
  2023PhRvX..13a1048A}. Based on Gravitational-Wave Transient Catalog
3 \citep[GWTC-3;][]{2021arXiv211103606T}, the primary BH mass
distribution ranges from $\sim 5\;\msun$ to $\sim 85\;\msun$ with the
global maximums of $\sim 10\;\msun$. The distribution does not
decrease monotonically from $\sim 10\;\msun$ to $\sim 85\;\msun$, and
has an overdensity at $\sim 35\;\msun$ and possible another
overdensity at $\sim 17\;\msun$.

In this paper, we mainly focus on merging binary BHs with primary BHs
of $\gtrsim 65\;\msun$. Such merging binary BHs were thought not to be
formed through isolated binary evolution because of PISNe as described
in section \ref{sec:PairInstability}. Thus, the discovery of GW190521
with $\sim 85\;\msun$ primary BH \citep{2020PhRvL.125j1102A,
  2020ApJ...900L..13A} seemed to favor the formation of merging binary
BHs in dense stellar clusters \citep{2019PhRvD.100d3027R,
  2020MNRAS.497.1043D, 2021ApJ...908..194T}. After the discovery,
several suggestions have been made that isolated binaries can also
form such binary BHs \citep{2020ApJ...905L..15B,
  2021MNRAS.501L..49K}. We have also proposed that Pop III stars (see
in section \ref{sec:PopIII}) can form GW190521-like binary BHs
\citep{2021MNRAS.505.2170T}. In this paper, we review these results
and related works.

\subsection{Pair instability supernovae}
\label{sec:PairInstability}

Stellar evolution theory predicts that very massive stars (say
$100$-$300\;\msun$ at the zero-age main-sequence times, ZAMS times for
short) cause thermonuclear explosions driven by PI, called ``PISNe''
\citep{1964ApJS....9..201F, 1967PhRvL..18..379B, 1968Ap&SS...2...96F,
  1983A&A...119...61O, 1984ApJ...280..825B, 1986A&A...167..274E,
  2001ApJ...550..372F, 2002ApJ...567..532H, 2002ApJ...565..385U,
  2012ARA&A..50..107L}. A PISN proceeds as follows. A very massive
star contains a carbon-oxygen (CO) core embedded in helium core and
hydrogen envelope several Myr after its formation. The CO core has
high temperature enough to cause pair production in which a photon is
converted into an electro-positron pair. The pair production decreases
pressure in the CO core radiation-dominated, and then the CO core
implodes. This implosion suddenly increases temperature in the CO
core, and triggers explosive carbon and oxygen burning. This explosive
burning finally overcomes the implosion, and conversely explodes the
very massive star. This explosion leaves behind no stellar
remnant. PISNe have not been conclusively discovered so far, although
there are several promising candidates \citep{2017NatAs...1..713T,
  2023arXiv230505796S}.

PISNe shape a mass range without BHs, so-called PI mass gap. This is
because PISNe leave behind no stellar remnant, and stars with ZAMS
masses of $\lesssim 100\;\msun$ and $\gtrsim 300\;\msun$ leave behind
BHs with quite different masses \citep{2002ApJ...567..532H,
  2017MNRAS.470.4739S}. A $\gtrsim 300\;\msun$ star experiences PI,
however fails to explode. The carbon and oxygen burning products are
photodissociated into hydrogen and helium before explosion. Since this
is endothermic reaction, the CO core pressure is decreased, and the
whole star collapses to a BH. The BH mass is at least $\sim
130\;\msun$, which corresponds to helium core mass when PI sets in. A
$\lesssim 100\;\msun$ star experiences PI, and partially explodes
because of weak carbon and oxygen burning. This explosion is so-called
pulsational pair instability supernovae (PPISNe)
\citep{2002ApJ...567..532H, 2007Natur.450..390W, 2012ARA&A..50..107L,
  2014ApJ...792...28C, 2016MNRAS.457..351Y, 2017ApJ...836..244W,
  2019ApJ...882...36M, 2019ApJ...887...53F, 2019ApJ...887...72L,
  2020A&A...640A..56R}. PPISNe leave behind $\sim 40\;\msun$
BHs. Further lower-mass stars experience core-collapse supernovae, and
leave behind $\lesssim 60\;\msun$ BHs for Pop I/II stars. Thus, PI
mass gap ranges from $\sim 60\;\msun$ to $\sim 130\;\msun$ for Pop
I/II stars.

The PI mass gap can deviate from the above range, and is especially
sensitive to the $\cago$ reaction rate \citep{2018ApJ...863..153T,
  2020ApJ...902L..36F, 2022MNRAS.516.1072C}. The $\cago$ reaction rate
controls the ratio of carbon to oxygen at the ending time of the core
helium burning phase; the carbon abundance becomes larger with the
$\cago$ reaction rate decreasing. After the core helium burning phase,
the higher carbon abundance makes carbon-carbon reactions more
active. The more active carbon-carbon reactions effectively reduce
helium core mass. Thus, the PI mass gap is shifted upward with smaller
$\cago$ reaction rate. If the $\cago$ reaction rate is $3\sigma$ (or 3
times) smaller than the standard value \citep{2013ApJS..207...18S},
the PI mass gap is shifted from $\sim 60$-$130\;\msun$ to $\sim
90$-$180\;\msun$. In that case, GW190521 can be formed through
isolated binary evolution. However, in this paper, we just adopt the
standard value unless stated.

\subsection{Population III stars}
\label{sec:PopIII}

Pop III stars, also known as first stars or metal-free stars, are made
of pristine gas at the high-redshift universe. Pop III stars have been
theoretically studied. They are typically formed as massive stars from
$10\;\msun$ to $1000\;\msun$ \citep{1998ApJ...508..141O,
  2002Sci...295...93A, 2004ARA&A..42...79B, 2008Sci...321..669Y,
  2011Sci...334.1250H, 2011MNRAS.413..543S, 2012MNRAS.422..290S,
  2013RPPh...76k2901B, 2013ApJ...773..185S, 2014ApJ...792...32S,
  2015MNRAS.448..568H, 2020ApJ...892L..14S}. They can have low-mass,
$\sim 1\;\msun$, as minor components \citep{2001ApJ...548...19N,
  2008ApJ...677..813M, 2011ApJ...727..110C, 2011Sci...331.1040C,
  2011ApJ...737...75G, 2012MNRAS.424..399G, 2013MNRAS.435.3283M,
  2014ApJ...792...32S, 2016MNRAS.463.2781C}. Massive Pop III stars
have not been directly observed so far \citep{2013MNRAS.429.3658R},
because they are located at the high-redshift universe due to their
short-lived nature. Low-mass Pop III stars have been discovered
neither \citep{2015ARA&A..53..631F, 2018MNRAS.473.5308M,
  2019MNRAS.487..486M}. Extremely metal-poor stars may be Pop III
stars polluted by interstellar medium
\citep{2015ApJ...808L..47K}. However, several studies have pointed out
that Pop III stars cannot be polluted up to the level of extremely
metal-poor stars \citep{2017ApJ...844..137T, 2018PASJ...70...34S,
  2018PASJ...70...80T,2019MNRAS.486.5917K}.

It is under debate if Pop III binary stars contribute to merging
binary BHs observed. Several studies have suggested that Pop III
binary stars little dominate merging binary BHs
\citep{2016MNRAS.460L..74H, 2017MNRAS.471.4702B}. On the other hand,
it has been argued that Pop III binary stars shape the overdensity at
$\sim 35\;\msun$ in the primary BH mass distribution
\citep{2021MNRAS.504L..28K}. Pop III binary stars may dominate merging
binary BHs in the PI mass gap \citep{2021MNRAS.505.2170T,
  2022ApJ...926...83T}, which is the main topic of this paper.

We briefly explain how Pop III binary stars fill the PI mass gap. Pop
III stars are expected to lose little mass through stellar winds,
since stellar winds becomes weaker with metallicity smaller
\citep{2005A&A...442..587V}. A Pop III star with $60$-$100\;\msun$ can
keep its mass until it ends its life. It forms a CO core not large
enough to cause PI, and then produces a iron core. The iron core
collapses at some point. The Pop III star leaves behind
$60$-$100\;\msun$ BHs without mass ejection. Then, we can find that
Pop III stars can fill the lower side of the PI mass gap. In section
\ref{sec:PopulationIIIBinaryBlackHoles}, we will assess if such BHs
can be genuinely members of merging binary BHs.

\subsection{Short summary}
\label{sec:ShortSummary}

In the above sections, we overview merging binary BHs, in particular,
those with BHs in the PI mass gap. Such merging binary BHs are
suggested to be possibly formed from Pop III binary stars. In sections
\ref{sec:BinaryPopulationSynthesisSimulation} and \ref{sec:Results},
we describe the method to confirm the suggestion, and its results,
respectively. In section \ref{sec:Summary}, we summarize this paper.

\section{Binary population synthesis simulation}
\label{sec:BinaryPopulationSynthesisSimulation}

We describe the method to investigate isolated binary evolution. In
section \ref{sec:InitialConditions}, we show initial conditions we
adopt. In section \ref{sec:SingleAndBinaryEvolutionModel}, we overview
single and binary evolution models.

\subsection{Initial conditions}
\label{sec:InitialConditions}

\begin{figure}[b]
\includegraphics[width=0.5\textwidth]{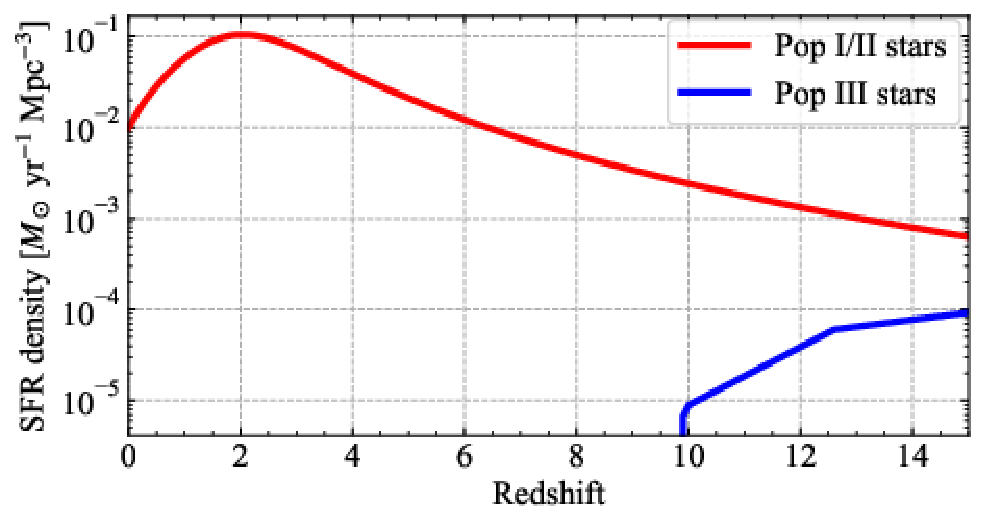}
\caption{\label{fig:sfrd} Redshift evolution of Pop I/II and Pop III
  SFR density.}
\end{figure}

We take into account the redshift evolutions of the star formation
rate (SFR) densities for Pop I/II and Pop III stars separately. Figure
\ref{fig:sfrd} shows Pop I/II and Pop III SFR density evolution as a
function of redshift. We adopt the formula of
\cite{2017ApJ...840...39M} for the Pop I/II SFR density
evolution. Note that the Pop I/II SFR density is updated by recent
observations, especially for high redshift \citep{2022ApJS..259...20H,
  2023ApJS..265....5H}. On the other hand, we simplify the simulation
result of \cite{2020MNRAS.492.4386S} for the Pop III SFR density
evolution.

Here, we note Pop III SFR density evolution we adopt. The simulation
of \cite{2020MNRAS.492.4386S} has considered Pop III star formation
and feedback, and followed the ending of Pop III star formation due to
metal enrichment. We construct our Pop III SFR density evolution to
approximate the Pop III SFR density evolution in fig. 4 of
\cite{2020MNRAS.492.4386S} by connecting three line segments shown in
eq. (5) in \cite{2022ApJ...926...83T}.  We use a Pop III SFR density
evolution based on a theoretical study, because Pop III stars have not
yet been observed unlike Pop I/II stars. We note that other studies
have also suggested Pop III SFR density evolution
\citep{2019MNRAS.488.2202J, 2020MNRAS.495.2475L, 2020ApJ...897...95V,
  2022ApJ...936...45H}, and some studies have constrained the upper
limit of Pop III SFR density evolution from observations
\citep{2011A&A...533A..32D, 2016MNRAS.461.2722I}. Our model may stop
Pop III star formation earlier than other models, for example,
suggested by \cite{2020ApJ...897...95V} who have shown that
reionization makes Pop III star formation long-lived.

\begin{figure}[b]
\includegraphics[width=0.5\textwidth]{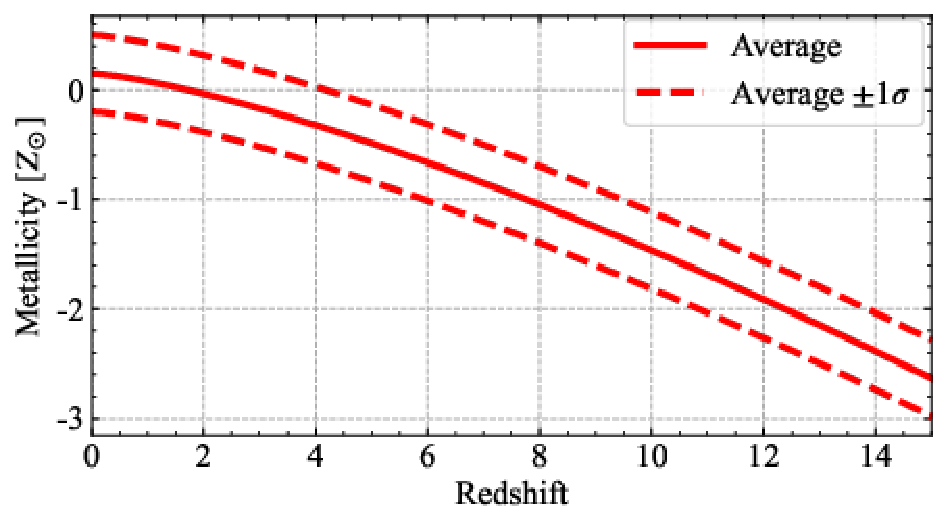}
\caption{\label{fig:avemetal} Redshift evolution of the average
  metallicity of Pop I/II stars (solid curve). We also show $\pm
  1\sigma$ deviations of the metallicity distribution for each
  redshift (dashed curves).}
\end{figure}

Figure \ref{fig:avemetal} shows the average metallicity of Pop I/II
stars formed at each redshift. For each redshift, stellar
metallicities are distributed as a logarithmic normal distribution
centered on the average metallicity with dispersion of $0.35$ dex in
logarithmic scale. We base the average metallicity evolution on the
analysis of observational data by \cite{2017ApJ...840...39M}, and
simplify the metallicity dispersion referring to the simulation
results of \cite{2019MNRAS.488.5300C}.  Figure \ref{fig:avemetal} also
indicates metallicities lower and higher than the average one by
$1\sigma$ for each redshift. For Pop III stars, we assume that all of
them have zero metallicity.

\begin{figure}[b]
\includegraphics[width=0.5\textwidth]{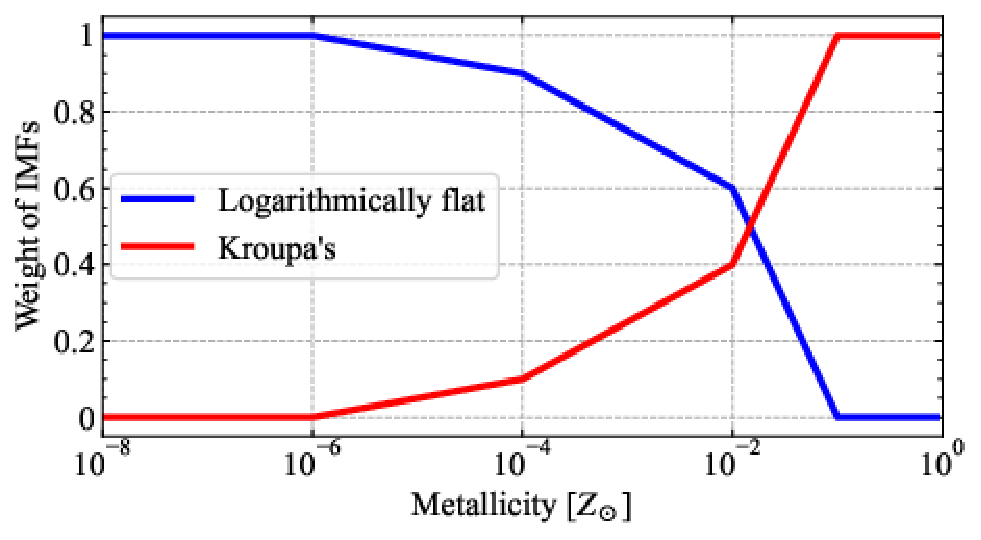}
\caption{\label{fig:imfweight} Dependence of IMF weights on
  metallicity. The red and blue curves indicate Kroupa's and
  logarithmically flat IMFs, respectively.}
\end{figure}

The initial conditions of single and binary stars are as follows. We
assume that the binary fraction is $0.5$ in number for all
metallicities. This binary fraction is similar to an intrinsic binary
fraction obtained by \cite{2012Sci...337..444S}. The initial mass
function of single stars and primary stars is set as a combination of
two mass functions. The first mass function is Kroupa's IMF
\citep{2001MNRAS.322..231K}. The minimum and maximum masses are $0.08$
and $150 \msun$, respectively. The second one is a function flat in
the logarithmic scale (hereafter, logarithmically flat IMF) from
$10\;\msun$ to $150\;\msun$, which is modeled as Pop III IMF
\citep{2014ApJ...781...60H, 2014ApJ...792...32S,
  2015MNRAS.448..568H}. We change weights of the two IMFs depending on
metallicities, which is motivated by star formation simulations with
various metallicities \citep{2021MNRAS.508.4175C,
  2022MNRAS.514.4639C}. We show the weights in Figure
\ref{fig:imfweight}. The weights of Kroupa's IMF are unity and zero
for metallicity of $\ge 0.1\;\zsun$ and $\le 10^{-6}\;\zsun$,
respectively, where we assume $\zsun = 0.02$. Between these
metallicities, the weights decreases gradually. We adopt Sana's model
\citep{2012Sci...337..444S} for binary initial conditions: the
distributions of mass ratios, periods, and orbital eccentricities.

Similarly to our Pop III SFR density evolution, we need to rely on
theoretical studies for determining Pop III IMF. We construct our Pop
III IMF, referring to fig. 9 in \cite{2014ApJ...781...60H} and fig. 6
in \cite{2015MNRAS.448..568H}. Our logarithmically flat IMF does not
deviate from their results by an order of magnitude.

\subsection{Single and binary evolution model}
\label{sec:SingleAndBinaryEvolutionModel}

We investigate the formation of merging binary BHs through isolated
binary evolution, employing rapid binary population synthesis
simulation with the BSEEMP code \citep{2020MNRAS.495.4170T,
  2022ApJ...926...83T}, which is based on the BSE code
\citep{2000MNRAS.315..543H, 2002MNRAS.329..897H}. Note that the BSE
code simultaneously solves single star evolution from ZAMS stars to
stellar remnants (e.g. white dwarfs, neutron stars, and BHs) with
stellar wind mass loss, and binary star evolution, such as wind
accretion, tidal interaction, stellar mass transfer, common envelope
evolution, and orbital decay due to gravitational wave radiation. The
BSEEMP code adopts different single and binary evolution model from
the BSE code. Here, we overview the single and binary evolution model
adopted by the BSEEMP code. In particular, we concentrate on different
parts between the BSE and BSEEMP codes.

\begin{figure}[b]
\includegraphics[width=0.5\textwidth]{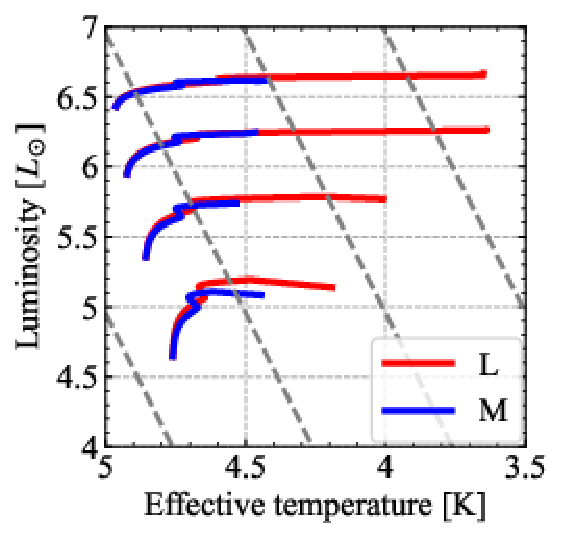}
\caption{\label{fig:hrdpop3} Hertzsprung-Russell diagram of M- and
  L-model stars with $Z=10^{-8}\;\zsun$ equivalent to Pop III
  stars. Their masses are $20$, $40$, $80$, and $160\;\msun$ from
  bottom to top. The dashed lines indicate stellar radii of $1$, $10$,
  $10^2$, and $10^3\;\rsun$ from left to right.}
\end{figure}

In the BSEEMP code, single stars evolve along with fitting formulae of
their total masses, helium core masses, radii, and luminocities. Note
that the BSE-based code like the BSEEMP code can rapidly follow single
and binary star evolution, since such fitting formulae are prepared in
advance. In the BSEEMP code, single stars with metallicities of
$>0.1\;\zsun$ evolve the same as in the BSE code. Note that the single
star evolution model has been constructed by
\cite{1998MNRAS.298..525P}. On the other hand, single stars with
metallicities of $\le 0.1\;\zsun$ evolve along with different fitting
formulae from the BSE code. Additionally, as one of the important
features of the BSEEMP code, the BSEEMP code supports $Z \le 5 \times
10^{-3} \zsun$, while the BSE code does not. The BSEEMP code prepares
two types of fitting formulae called ``M model'' and ``L model''
constructed in \cite{Yoshida19} with the HOSHI code
\citep{2016MNRAS.456.1320T, 2018ApJ...857..111T, Takahashi19}. The
``M'' and ``L'' stand for ``M''ilky Way and ``L''arge Magellanic
Cloud, respectively. \cite{Yoshida19} have referred stars in these
galaxies when they construct these models. The difference between the
M and L models is the efficiency of overshoot at the convective
boundary, and the overshoot in the M model is smaller than in the L
model as summarized in Appendix A of \cite{2022ApJ...926...83T}. For
extreme metal-poor cases, radius evolutions of M- and L-model stars
are quite different. Figure \ref{fig:hrdpop3} shows the
Hertzsprung-Russell diagram of M- and L-model stars with
$Z=10^{-8}\;\zsun$ equivalent to Pop III stars. M-model stars expand
up to $\sim 100\;\rsun$ even for a $160\;\msun$ star, while L-model
stars exceed $10^3\;\rsun$ for $80$ and $160\;\msun$ stars. If we
switch off the overshooting effect, the maximum radii should be
smaller than those of M-model stars.

The reason for the difference of the maximum radii between M- and L-
models is as follows. A L-model star's core gains fresh hydrogen from
its envelope due to more efficient overshoot during its main-sequence
(MS) phase. Eventually, a L-model star has a larger helium core mass
than a M-model star at the ending of their MS phases by $20$ \%. As a
result of this mass difference, a L-model star has much larger core
luminosity than a M-model star at their post-MS (PMS) phases. Since
larger core luminosity makes stellar radius larger, a L-model star has
a much larger maximum radius than a M-model star.

Here, we compare the M- and L-models with other models. The radius
evolution of a L-model star is similar to the Pop III model by
\cite{2001A&A...371..152M}. Stars with $\gtrsim 80 \msun$ finally
evolve to red supergiant stars in both the Marigo's and L models. On
the other hand, M-model stars evolve like the Pop III model by
\cite{2021MNRAS.502L..40F}. Even stars with $\sim 80 \msun$ keep blue
supergiant stars throughout their lives in both the Farrell's and M
models.

Here, we describe how to separate four stellar phases: MS,
Hertzsprung-gap, core helium burning, and shell helium burning
phases. When stars run out of hydrogen at their core, the whole stars
shrink, and their effective temperature once become larger. We define
the ending of the MS phase as the time when the effective temperature
gets the local maximum, which can be seen at the effective temperature
of $\sim 10^{4.7}$ K in Figure \ref{fig:hrdpop3}. The ending of
Hertzsprung-gap, core helium burning, and shell helium burning phases
can not be seen in Hertzsprung-Russell diagram like Figure
\ref{fig:hrdpop3}. We read required information from the results of
the HOSHI code when we construct our fitting formulae. We define the
ending of the Hertzsprung-gap phase as the time when helium is ignited
at the stellar core, or the central temperature exceeds $\sim 2 \times
10^8$ K. We define the ending of the core helium burning phase as the
time when the central helium fraction becomes sufficiently small. We
set the fraction to $0.1$ \%. We define the ending of the helium shell
burning phase as the time when carbon is ignited at the stellar core,
or the central temperature exceeds $\sim 10^9$ K. Note that Pop III
stars have no Hertzsprung phase unlike Pop I/II stars. They have
enough high central temperature to ignite helium at the ending of
their MS phases. Thus, Pop III stars skip Hertzsprung-gap phases in
the BSEEMP code.

\begin{figure}[b]
\includegraphics[width=0.5\textwidth]{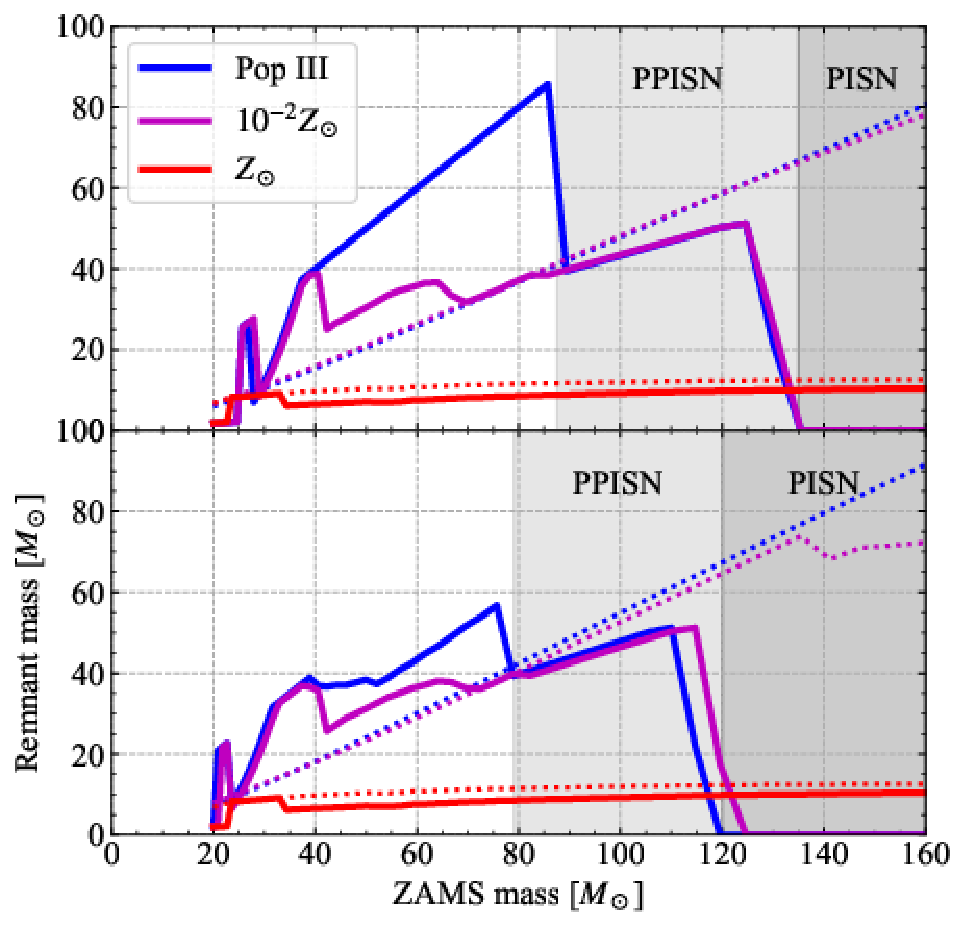}
\caption{\label{fig:zamsremnant} Relation between ZAMS and remnant
  masses (solid) and between ZAMS and helium core mass just before
  core collapse (dotted) for $Z=\zsun$, $Z=10^{-2}\zsun$ and Pop
  III stars. The top and bottom panels show the M and L models,
  respectively.  Light-gray and dark-gray regions indicate those in
  which PPISN and PISN are effective, respectively, for
  $Z=10^{-2}\zsun$ and Pop III stars. $Z=\zsun$ stars do not
  experience PPISN nor PISN because of their small helium cores just
  before core collapse.}
%%We adopt the M model for this plot, and obtain a similar plot even
%%if we adopt the L model.}
\end{figure}

Single star evolutions along with the fitting formulae are modified by
stellar wind mass loss. The BSEEMP code adopts Belczynski's stellar
wind model \citep{2010ApJ...714.1217B}, which includes OB star winds
\citep{Vink01}, giant branch star winds \citep{1978A&A....70..227K,
  1983ARA&A..21..271I}, Wolf-Rayet star winds
\citep{1998A&A...335.1003H, 2005A&A...442..587V}, asymptotic giant
branch star winds \citep{1993ApJ...413..641V}, and luminous blue
variable star winds \citep{1994PASP..106.1025H}. The OB and Wolf-Rayet
star winds are metallicity-dependent, such that they becomes weak with
metallicity decreasing.

Massive stars ($\gtrsim 8\;\msun$) finally experience supernovae, and
leave behind neutron stars, BHs, or no stellar remnant. Supernova
model in the BSEEMP code is the Fryer's rapid model
\citep{2012ApJ...749...91F} with the modification of PI mass loss
modeled as the moderate PPISN model \citep{2019ApJ...887...72L,
  2020A&A...640L..20B}. Note that several PPISN models have been
suggested \citep{2019ApJ...882..121S, 2020ApJ...904L..13R}.

We evolve $20$-$160 \msun$ stars from their ZAMS phases to their
remnant phases, taking into account stellar wind mass loss and
supernova mass loss. We can see the effects of stellar winds and
supernovae in Figure \ref{fig:zamsremnant}. Remnant masses increase
with metallicities decreasing because of weakening stellar winds. Pop
III remnant masses are suddenly reduced at ZAMS masses of $\sim
80$-$85\msun$ and $\sim 120$-$135\msun$ due to PPISN and PISN effects,
respectively. Note that helium core masses just before core collapse
exceed $\sim 40 \msun$ and $\sim 60 \msun$ at ZAMS masses of $\sim
80$-$85\msun$ and $\sim 120$-$135\msun$, respectively. PPISN and PISN
can equalize remnant masses of $Z=10^{-2}\zsun$ and Pop III
stars. Especially for the M model, Pop III stars with ZAMS masses of
$60$ -- $85\msun$ can fill the lower side of the PI mass gap. This is
because they keep their initial mass before core collapse, and do not
experience PPISN nor PISN.

The ZAMS-remnant mass relations are similar between the M and L
models, however slightly different. ZAMS mass ranges for PPISNe and
PISNe are shifted downward for the L model. This is because L-model
stars create larger helium core masses due to more efficient overshoot
than M-model stars during their MS phases. This effect is noticeable
in the difference of the maximum BH mass between the L and M models
($\sim 60 \msun$ and $\sim 85 \msun$, respectively).

Neutron stars and BHs should receive natal kicks due to asymmetric
supernova explosions following core collapses. We assume this natal
kicks, such that the velocity distribution is given by a single
Maxwellian with 265 km~s$^{-1}$ \citep{2005MNRAS.360..974H} if there
is no fallback mass. We reduce the natal kick velocity by $1-f_{\rm
  b}$ where $f_{\rm b}$ is the fraction of the fallback mass
\citep{2012ApJ...749...91F}. We suppose that the directions of the
natal kicks are isotropic.

We construct our binary evolution model, the BSEEMP model, based on
the BSE model \citep{2002MNRAS.329..897H} with several
modifications. The BSE model considers wind accretion, tidal
evolution, stable mass transfer, common envelope evolution, magnetic
braking, and orbital decay through gravitation wave radiation. We
modify prescriptions of tidal evolution, stable mass transfer, and
common envelope evolution. Here, we focus on different points of these
processes from the BSE model.

A star of a binary star evolves to a giant star at some point, and
fills its Roche lobe unless the orbit of the binary star is too a
wide. The star (hereafter, donor) transfers its mass toward its
companion star (hereafter accretor). If the mass transfer is stable,
the binary star experiences stable mass transfer. Otherwise, the
binary star experiences stellar coalescence, or common envelope
evolution, depending on their stellar types as described in detail
below. A donor can sustain stable mass transfer if it has a radiative
envelope, whereas a donor with a convective envelope can cause
unstable transfer.  This is because a star with a radiative envelope
more sensitively decreases its radius with decreasing its mass. We
modify the boundary criteria between radiative and convective
envelopes. In the BSE model, a massive star changes its envelope from
radiative to convective when it finishes core helium burning, and
starts shell helium burning. On the other hand, in the BSEEMP model, a
massive star has a radiative (convective) envelope when the effective
temperature of its envelope is above (below) $10^{3.65}$~K. In
general, whether an envelope is radiative or convective depends on the
effective temperature. Moreover, in the M and L models, a massive star
can start shell helium burning at an effective temperature much higher
than $10^{3.65}$~K, especially for metal-poor stars, such as Pop III
stars. Thus, this criteria are very reasonable. Similar criteria have
been suggested \citep{2014MNRAS.442.2963K, 2021A&A...651A.100O}.

We also modify the prescription of stable mass transfer. In the BSE
model, all the mass lost by a donor are transferred onto an accretor
unless the mass accretion rate exceeds Eddington accretion limit. In
the BSEEMP model, we can control the fraction of transferred
mass. Here, we fiducially set the fraction to $0.5$. This choice can
be seen in several rapid binary population synthesis codes
\citep{2020A&A...636A.104B, 2020MNRAS.498.3946K}. Since the fraction
can be not well-established, and range from zero to unity
\citep{1989A&A...226...88M}, we take the middle. This is consistent
with a recent work that the fraction should be $\lesssim 0.5$
\citep{2017NatCo...814906S}.

In the BSE model, if post main-sequence stars including
Hertzsprung-gap, core helium burning, and shell helium burning stars
fills its Roche lobe, and its mass transfer is unstable, a binary star
experiences common envelope evolution. However, in the BSEEMP model,
this is true only for core helium burning and shell helium burning
stars. If a Hertzsprung-gap star fills its Roche lobe, and its mass
transfer is unstable, the binary star experience stellar
coalescence. This is because a Hertzsprung-gap star does not yet
develop steep density gradient between the helium core and hydrogen
envelope \citep{2004ApJ...601.1058I}. Such a prescription is also
equipped with several rapid binary population synthesis codes
\citep{2012ApJ...759...52D, 2018MNRAS.474.2959G}.

Pop III stars, even at the ZAMS time, the core region contracts
sufficiently so that the central temperature reaches $\sim 10^8$ K,
where carbon is produced by helium burning. Thus, hydrogen burning
stably proceeds via the CNO cycle. Nevertheless, Pop III stars do not
have steep density gradient between their core and hydrogen envelope
in their MS phases. This is because Pop III stars contracts not only
core regions but also overall regions at the ZAMS time.

In common envelope evolution, a binary orbit shrinks as a back
reaction of common envelope ejection. In rapid binary population
synthesis codes, such a common envelope evolution is modeled as
so-called the $\alpha$ formalism \citep{1984ApJ...277..355W}. In this
formalism, two parameters $\alpha$ and $\lambda$ have to be set. A
parameter $\alpha$ controls the efficiency at which a binary orbital
binding energy is converted to energy driving common envelope ejection
through gas drag. We set $\alpha=1$. A parameter $\lambda$ is a
dimensionless parameter of the binding energy of a donor's envelope or
common envelope. We adopt the formulae of \cite{2014A&A...563A..83C}
for $\lambda$.

We adopt a new prescription for tidal evolution of stars with
radiative envelopes. The tidal evolution follows the radiative damping
of the dynamical tide \citep{1975A&A....41..329Z}. In the BSE model,
the tidal coupling parameter is given by \cite{1977A&A....57..383Z}
and \cite{1981A&A....99..126H}. We replace this tidal coupling
parameter with a new formula derived by \cite{2010ApJ...725..940Y} and
\cite{2018A&A...616A..28Q}. This formula is also used by
\cite{2020MNRAS.498.3946K}.

\subsection{Realization}
\label{sec:Realization}

We prepare two parameter sets which adopt the M and L models. We
consider a parameter set with the M model as the fiducial model, and
call it ``fid'' set. We call the other one the ``L'' set.

For each parameter set, we generate 12 groups of $3 \times 10^6$
binary stars. 8 groups of the 12 groups have Kroupa's IMF and
metallicities of $Z=\zsun, 0.5\zsun, 0.25\zsun, 0.1\zsun, 0.05\zsun,
0.025\zsun, 10^{-2}\zsun$, and $10^{-4}\zsun$. The rest of the 12
groups have the logarithmically flat IMF and metallicities of
$10^{-2}\zsun, 10^{-4}\zsun, 10^{-6}\zsun$, and $10^{-8}\zsun$.

We generate primary stars of binary stars, such that the minimum and
maximum masses are $10$ and $150 \msun$. Moreover, we set the lower
limit of the secondary stars to be $10 \msun$. This is because binary
stars with $<20 \msun$ in total can form no BH even if the binary
components merge.

\section{Results}
\label{sec:Results}

In this section, we show several features of our fid set. In sections
\ref{sec:PopulationIIIBinaryBlackHoles} and
\ref{sec:PairInstabilitySupernovaRate}, we present properties of
merging binary BH and PISNe, respectively.

\subsection{Population III binary black holes}
\label{sec:PopulationIIIBinaryBlackHoles}

\begin{figure}[b]
\includegraphics[width=0.5\textwidth]{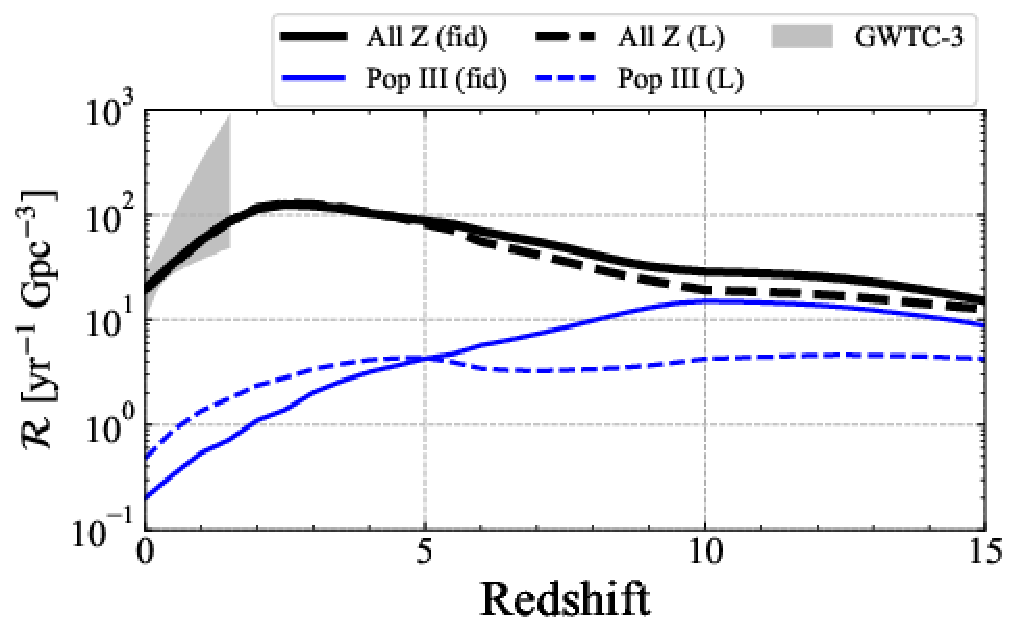}
\caption{\label{fig:mergerrate} Redshift evolution of binary BH merger
  rate densities for the fid and L sets, indicated by solid and dashed
  curves, respectively. Black and blue curves show the merger rate
  densities of all binary BHs and Pop III binary BHs,
  respectively. The gray shaded region indicates the corresponding
  merger rate density inferred by GWTC-3 within 90 \% credible
  bounds.}
\end{figure}

In this section, we compare our theoretical results with observed
binary BH population. First of all, we can see the redshift evolution
of binary BH merger rate densities for the fid and L sets in Figure
\ref{fig:mergerrate}. For both the fid and L sets, the merger rate
densities are $\sim 20$ yr$^{-1}$ Gpc$^{-3}$ at the present day. The
merger rate densities increase up to redshifts of 2 or 3, and decrease
beyond redshifts of 2 or 3. The features follow the redshift
evolution of SFR densities (see Figure \ref{fig:sfrd}). Note that the
merger rate densities are less sensitive to redshifts than SFR
densities. This is because binary BH mergers can be delayed by several
Gyr from star formation.

For the fid and L sets, the merger rate densities are quite consistent
with the GWTC-3 result. The merger rate densities are within 90 \%
credible bounds inferred by observations. Note that we obtain this
consistency because of prohibition of common envelope evolution for
Hertzsprung-gap stars (see section
\ref{sec:SingleAndBinaryEvolutionModel}). Unless we prohibit the
common envelope evolution of Hertzsprung-gap stars, we would obtain
$\sim 10$ times larger BH merger rate density than obtained in Figure
\ref{fig:mergerrate}. The merger rate densities are very sensitive to
binary evolution model.

We explain why the criteria of common envelope evolution affects the
merger rate density in more detail. It may be unclear why prohibition
of common envelope evolution for Hertzsprung-gap stars largely affects
the merger rate density, because Hertzsprung-gap stars cannot enter
into common envelope evolution due to their radiative
envelopes. Certainly, Hertzsprung-gap stars are not easy to enter into
common envelope evolution, but not impossible. If Hertzsprung-gap
stars have sufficiently larger masses than their companion stars, they
can enter into common envelop evolution. Furthermore, it may look
strange why M-model stars enter into common envelope evolution,
because Pop III stars always have radiative envelopes. Actually, Pop
II stars in the M model have convective envelopes just before core
collapse. Pop III stars in the M model can always have radiative
envelopes, not only because of their inefficient overshoot, but also
the small opacities of their metal-free envelopes. We can see
Hertzsprung-Russell diagrams of Pop II stars in the M model from
fig. 9 of \cite{2022ApJ...926...83T}.

At the present day, Pop III binary BHs contribute to only 1 and 3 \%
of the total merger rate densities for the fid and L sets,
respectively. They are always minor for the L set, while they become
dominant beyond a redshift of 10 for the fid set. The current
observatories detect binary BHs with non Pop III origins, although the
future observatories, such as Einstein Telescope
\citep{2010CQGra..27s4002P} and Cosmic Explorer
\citep{2019BAAS...51g..35R}, will discover many binary BHs originating
from Pop III stars.

\begin{figure}[b]
\includegraphics[width=0.5\textwidth]{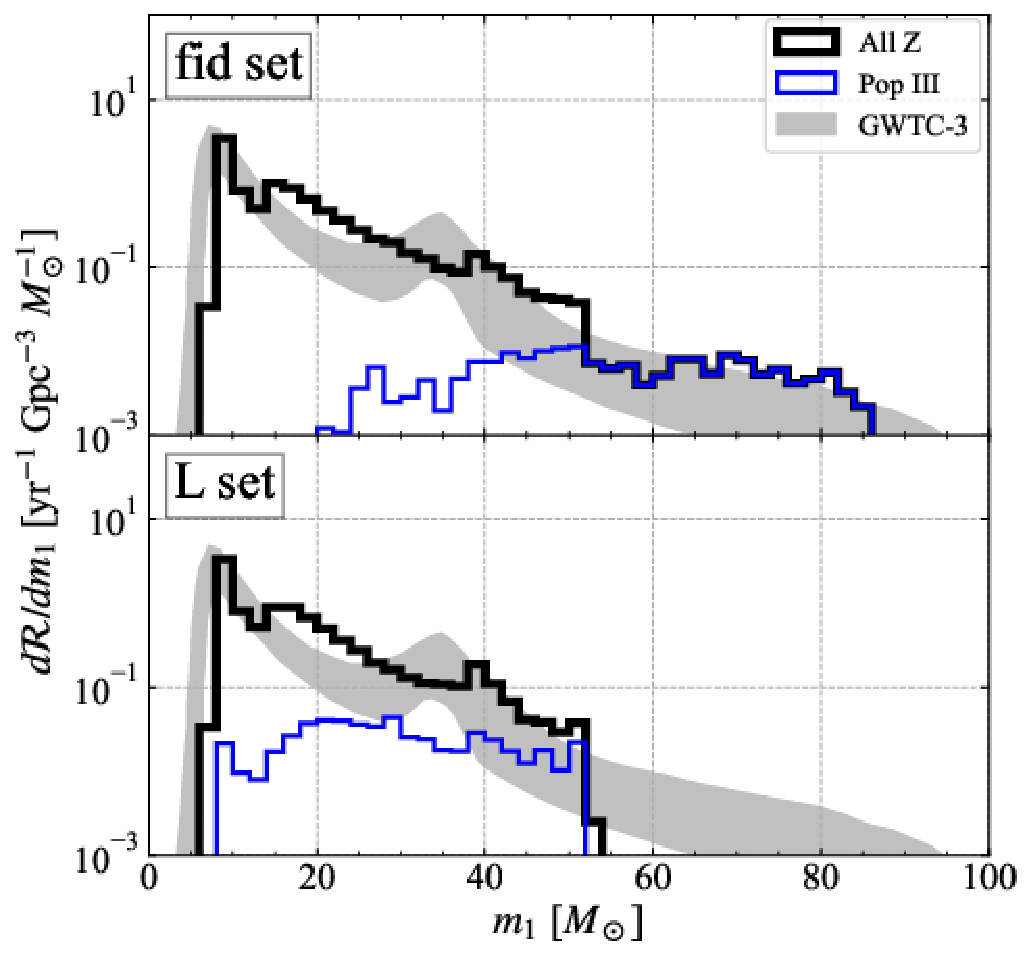}
\caption{\label{fig:primarymass} Primary BH mass distribution of
  merging binary BHs at a redshift of $0$ for the fid (top) and L
  (bottom) sets. Black and blue curves indicate all the binary BHs and
  Pop III binary BHs, respectively. The gray shaded region indicates
  the corresponding primary BH mass distribution inferred by GWTC-3
  within 90 \% credible bounds.}
\end{figure}

Figure \ref{fig:primarymass} shows the primary BH mass distributions
of merging binary BHs at the present day, where a primary BH mass (or
$m_1$) is the heavier BH mass of two BHs. For the fid set, the primary
BH mass ranges from $5\msun$ to $85\msun$, similarly to the GWTC-3
result. Moreover, Pop III binary BHs dominate binary BHs with $m_1
\gtrsim 50 \msun$. These BHs come from Pop III stars with their ZAMS
mass of $50$-$90 \msun$ (see Figure \ref{fig:zamsremnant}). In other
words, Pop III binary BHs fill the lower half of the PI mass gap ($60$
-- $130 \msun$). This feature is quite different from that of the L
set. For the L set, there is no binary BHs with $m_1 \gtrsim 55
\msun$. Even Pop III binary BHs cannot fill the PI mass gap at all.

As seen in Figure \ref{fig:primarymass}, there is overproduction at
$m_1 \sim 20\msun$ for both the fid and L sets. This comes from binary
BHs with Pop II origins. We may solve this overproduction,
constructing Pop I/II SFR density evolution more carefully
\citep{2023ApJ...948..105V}.

\begin{figure}[b]
  \includegraphics[width=0.48\textwidth]{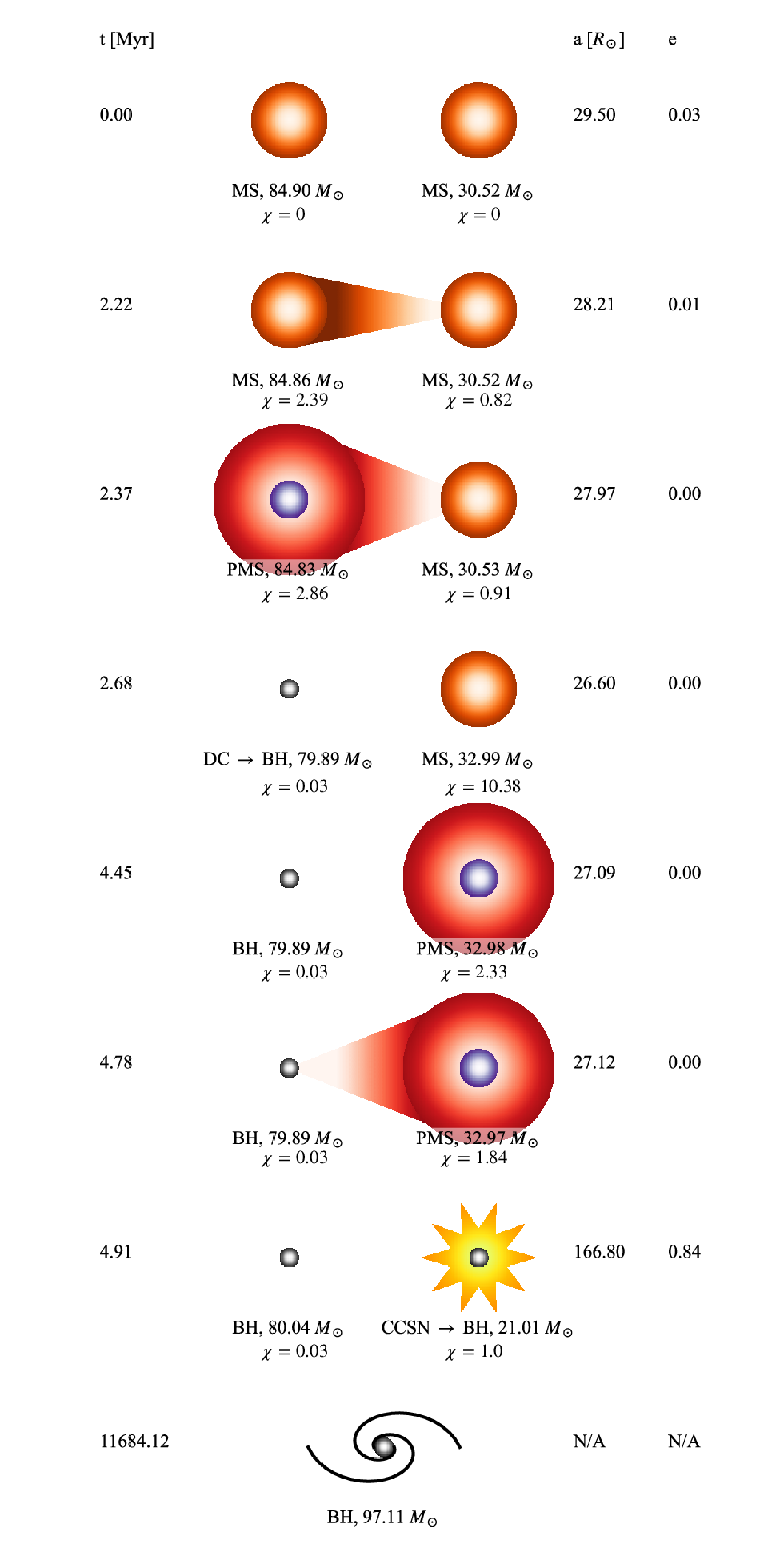}
\caption{\label{fig:fidset} Typical formation channel of a merging
  binary BH in the PI mass gap for the fid set. MS, PMS, and CCSN are
  short for main-sequence star, post main-sequence star, and core
  collapse supernova, respectively. The smaller spheres in PMSs
  indicate helium cores in PMSs. A dimensionless spin parameter
  ($\chi$) is attached to each star.}
\end{figure}

In Figure \ref{fig:fidset}, we draw the picture of a typical formation
channel of a merging binary BH in the PI mass gap for the fid set. We
prepare a Pop III binary star with the primary mass of $\sim 85
\msun$, the secondary mass of $\sim 31 \msun$, and the separation of
$\sim 30 \rsun$. The primary star evolves with little mass loss. Pop
III stars experience weak stellar wind mass loss. Moreover, for the
fid set, a $\sim 85 \msun$ star expands up to a few $10 \rsun$ (see
Figure \ref{fig:hrdpop3}), and loses little mass through stable mass
transfer. Eventually, the primary star leaves behind a BH with $\sim
80 \msun$ in the PI mass gap. The secondary star also experiences
little mass loss through stellar winds and stable mass
transfer. Because it experiences supernova mass loss, it leaves behind
a BH with $\sim 20 \msun$. Finally, they evolve to a binary BH with a
wide but eccentric orbit. Thus, they can merge within the Hubble time.

As seen in Figure \ref{fig:fidset}, progenitors of binary BHs in the
PI mass gap experience stable mass transfer. Nevertheless, their
formation channel does not depend on the choice of formulation of
stable mass transfer, such as the mass-transfer termination conditions
\citep{2017MNRAS.468.5020I}. This is because Pop III stars in the M
model experience little stable mass transfer due to their small
maximum radii.

In Figure \ref{fig:fidset}, we can see that the primary star loses
$0.07 \msun$ from $t=0$ Myr to $2.37$ Myr, while the secondary star
gains only $0.01 \msun$ during this time, despite that the fraction of
transferred mass is $0.5$. This is because the primary star its mass
mainly due to stellar wind mass loss, although it also experiences
mass transfer. This is the reason why the secondary star increases
only $0.01 \msun$.

For the L set, the channel in Figure \ref{fig:fidset} is
prohibited. Since the primary star expands beyond $10^3 \rsun$, it
loses a large amount of mass in its hydrogen envelope through stable
mass transfer or common envelope. Then, they leave behind $\sim 40
\msun$, not BHs in the PI mass gap. Instead of the above initial
condition, let's consider a binary star with the separation of
$\gtrsim 10^3 \rsun$. In that case, they do not experience mass loss
through binary interactions. Thus, they can leave $\sim 80 \msun$
BHs. However, their separation is still $\gtrsim 10^3 \rsun$. They
cannot merge within the Hubble time. This is the reason why there are
no merging BHs in $m_1 \gtrsim 55 \msun$.

From the above, we can conclude that Pop III stars can explain the
presence of merging binary BHs in the PI mass gap if their convective
overshoot is as inefficient as the M model. In the future, we need
whether the convective overshoot is efficient or not.

We mention other three features of the primary BH mass
distribution. First, for both the fid and L sets, the binary BH
population suddenly decreases below $m_1 \sim 8 \msun$. This is
because the supernova model we adopt tend not to form BHs with
$\gtrsim 8 \msun$. We note that this sudden decrease can be explained
not by the supernova model but by binary interaction
\citep{2022ApJ...940..184V}. Second, there is the overdensity at $\sim
35 \msun$ in the GWTC-3 result. However, our results cannot form the
overdensity. Although PPISNe are expected to form the overdensity they
fail to form \citep{2022RNAAS...6...25R, 2022ApJ...931...17V}. We may
not fully explain the primary BH mass distribution, although we can
form merging binary BHs in the PI mass gap. Third, there may be
another overdensity at $\sim 17\;\msun$ although the overdensity
cannot be seen in the GWTC-3 result of Figure
\ref{fig:primarymass}. It seems that the fid set has an overdensity at
$\sim 17\;\msun$. We have no intention of claiming that the
overdensity in the fid set is consistent with the overdensity at $\sim
17 \msun$ in the GWTC-3 result.

\begin{figure}[b]
\includegraphics[width=0.5\textwidth]{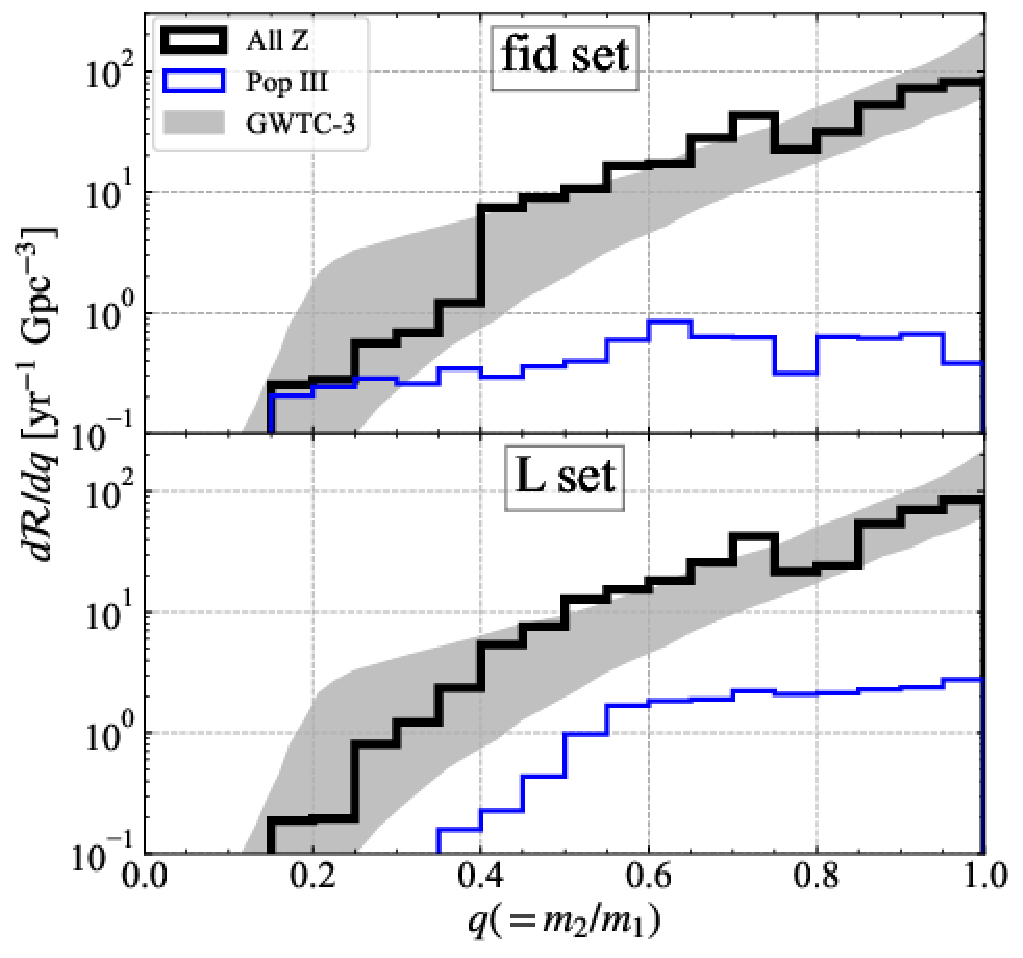}
\caption{\label{fig:massratio} Mass ratio distribution of merging
  binary BHs at the present day for the fid (top) and L (bottom)
  sets. Black and blue curves indicate all the binary BHs and Pop III
  binary BHs, respectively. The gray shaded region indicates the
  corresponding mass ratio distribution inferred by GWTC-3 within 90
  \% credible bounds.}
\end{figure}

We can see the mass ratio distribution of merging binary BHs at the
present day in Figure \ref{fig:massratio}. The mass ratio is defined
as the ratio of the secondary BH mass ($m_2$) to the primary BH mass,
where the secondary BH mass is the lighter BH mass of two BHs. As seen
in Figure \ref{fig:massratio}, the mass ratio distributions in our
results are within 90 \% credible bounds of the GWTC-3. Especially for
the fid set, binary BHs with low mass ratios ($\sim 0.2$) are formed
from Pop III binary stars with low mass ratios at the initial time. As
described above, Pop III binary stars leave behind merging binary BHs
with little mass loss through binary interaction. They can keep the
initial mass ratio. If Pop III binary stars have a mass ratio
distribution similar to Pop I/II binary stars, they can form merging
binary BHs with such low mass ratios.

\begin{figure}[b]
\includegraphics[width=0.5\textwidth]{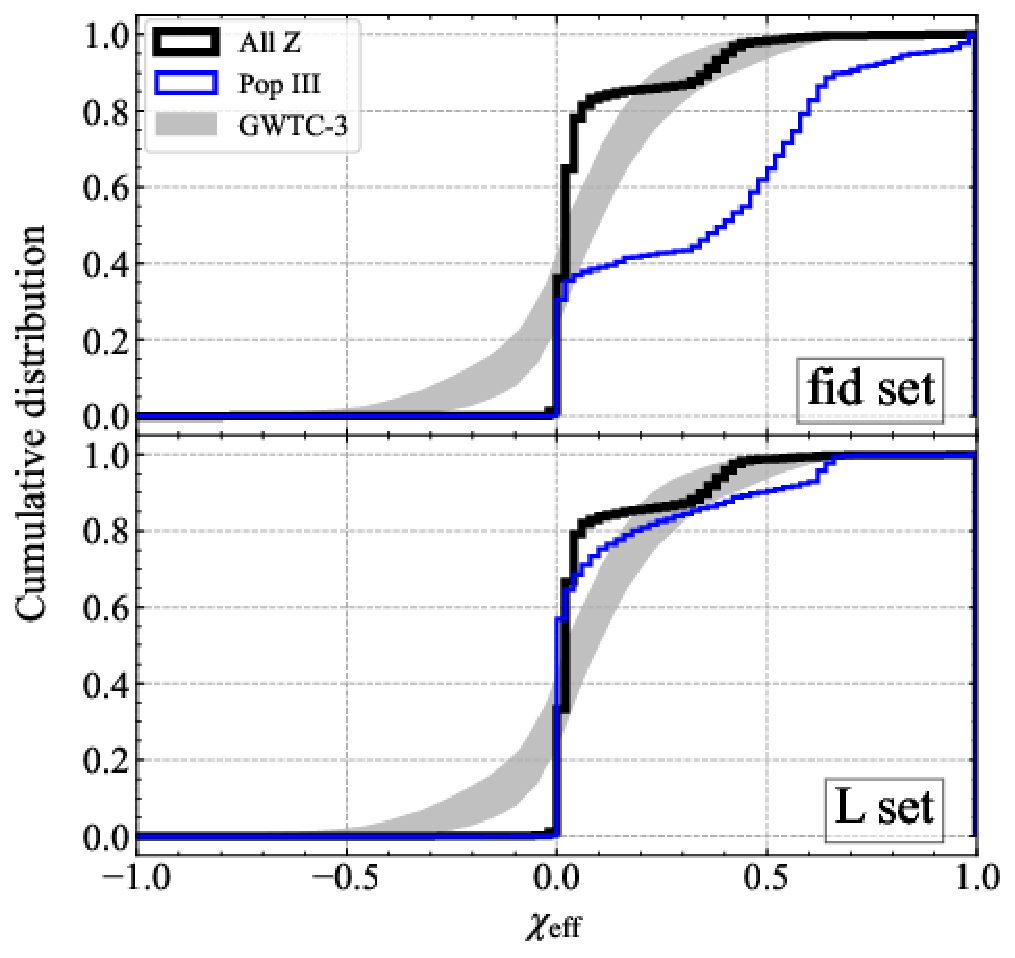}
\caption{\label{fig:effectivespin} Cumulative distribution of
  effective spins of merging binary BHs at the present day for the fid
  (top) and L (bottom) sets. Black and blue curves indicate all the
  binary BHs and Pop III binary BHs, respectively. The gray shaded
  region indicates the corresponding the cumulative distribution of
  effective spins inferred by GWTC-3 within 90 \% credible bounds.}
\end{figure}

Figure \ref{fig:effectivespin} shows the cumulative distribution of
effective spins of merging binary BHs at the present day for the fid
and L sets. An effective spin can be expressed as
\begin{align}
  \chi_{\rm eff} = \left( \frac{m_1 \overrightarrow{\chi}_1 + m_2
    \overrightarrow{\chi}_2}{m_1+m_2} \right) \cdot
  \frac{\overrightarrow{L}}{| \overrightarrow{L} |},
\end{align}
where $\overrightarrow{\chi}_1$ and $\overrightarrow{\chi}_2$ are,
respectively, dimensionless spin vectors of the primary and secondary
BHs, and $\overrightarrow{L}$ is the orbital angular momentum of a
binary BH. Roughly speaking, an effective spin indicates BH spin
component parallel to the orbital angular momentum of a binary BH. For
both the fid and L sets, most of merging binary BHs have positive
$\chi_{\rm eff}$, about 80 \% of them have $\chi_{\rm eff} \lesssim
0.1$, and the rest of them have $\chi_{\rm eff} \gtrsim 0.1$.

Merging binary BHs with $\chi_{\rm eff} \gtrsim 0.1$ have highly
spinning secondary BHs. Such BHs are mainly formed as follows
\citep{2016MNRAS.462..844K, 2017ApJ...842..111H}. Let's consider a
binary with one BH and one main-sequence star, where the BH mass is
smaller than the main-sequence star. Such a binary can appear after
the primary star collapses to the BH. When the main-sequence star
evolves off to a giant star, it fills its Roche lobe, and then starts
mass transfer. Regardless of the mass transfer stability, the binary
separation becomes smaller with the giant star losing its mass. The
giant star completely loses its hydrogen envelope, and evolves to a
naked helium star. The binary interaction shrinks the binary
separation down to $\sim 10 \rsun$. Then, the naked helium star is
tidally spun-up. Finally, the naked helium star collapses to a highly
spinning BH. When the binary BH merges, we can observe it as a binary
BH with high $\chi_{\rm eff}$.

As seen in Figure \ref{fig:effectivespin}, about 80 \% of merging
binary BHs have $\chi_{\rm eff} \lesssim 0.1$. They have large
separation (say $\gtrsim 10 \rsun$) when the giant star evolves to the
naked helium star, losing its hydrogen envelope completely. The naked
helium star cannot be tidally spun-up when the binary separation is
large. Eventually, binary BHs have larger effective spins when their
progenitors have smaller separation after common envelope
evolution. Initial separations of binary BH progenitors indirectly
affect effective spins of binary BHs, since binary BH progenitors have
smaller separations after common envelope evolution when their initial
separations are smaller.

As described above, tidal interaction between naked a helium star and
a BH is the primary source of high spins. This is the case for Pop
I/II binary BHs. In the case of Pop III binary BHs, the primary source
of high spins is stable mass transfer. As seen in Figure
\ref{fig:effectivespin}, a large fraction of Pop III binary BHs have
$\chi_{\rm eff} \gtrsim 0.4$ due to spin up in stable mass
transfer. This can be also seen in Figure \ref{fig:fidset}. However,
we emphasize that this spin-up process is not dominant for overall
binary BHs, since Pop III binary BHs are minor compared to Pop I/II
binary BHs.

As seen in Figure \ref{fig:mergerrate}, Pop III binary BHs have small
contribution. Thus, their effective spin distribution does not affect
the overall distribution. However, we explain how to form their
effective spin distribution especially for the fid set, because their
distribution is completely different from the overall
distribution. About 40 \% of Pop III binary BHs have $\chi_{\rm eff}
\gtrsim 0.1$. The reason is as follows. They can keep their hydrogen
envelope without stable mass transfer nor common envelope evolution
because of their small radii (see Figure \ref{fig:hrdpop3}). Thus,
they can keep their spin angular momenta throughout their
evolution. Finally, they can collapse to highly spinning BHs, and
merge as binary BHs with high $\chi_{\rm eff}$.

For the L set, Pop III binary BHs have an effective spin distribution
similar to the overall distribution. Pop III stars expand similarly to
Pop I/II stars. Thus Pop III BHs obtain spin angular momenta in the
same process as Pop I/II BHs: tidal spin-up during naked helium stars
as described above.

For both the fid and L sets, the cumulative distributions appear
different from the GWTC-3 result. In particular, our results contain
no merging binary BHs with negative $\chi_{\rm eff}$ in contrast to
the GWTC-3 result. However, this may not be much of a problem. Several
studies have suggested that binary BH population with positive
$\chi_{\rm eff}$ is real, but BH population with negative $\chi_{\rm
  eff}$ is spurious \citep{2021ApJ...921L..15G,
  2021PhRvD.104h3010R}. If this is true, our results are consistent
with the GWTC-3 result.

\subsection{Pair instability supernova rate}
\label{sec:PairInstabilitySupernovaRate}

We investigate features of PISNe for the fid set in order to forecast
current and future PISN surveys \citep{2012MNRAS.422.2701P,
  2019PASJ...71...59M, 2019PASJ...71...60W, 2020ApJ...894...94R,
  2022ApJ...925..211M, 2022A&A...666A.157M}. As reference, we prepare
a different parameter set which can produce merging binary BHs in the
PI mass gap despite of choice for the L model. In this section, we
first introduce the parameter set, called the ``L-$\tsig$'' set.

\begin{figure}[b]
\includegraphics[width=0.5\textwidth]{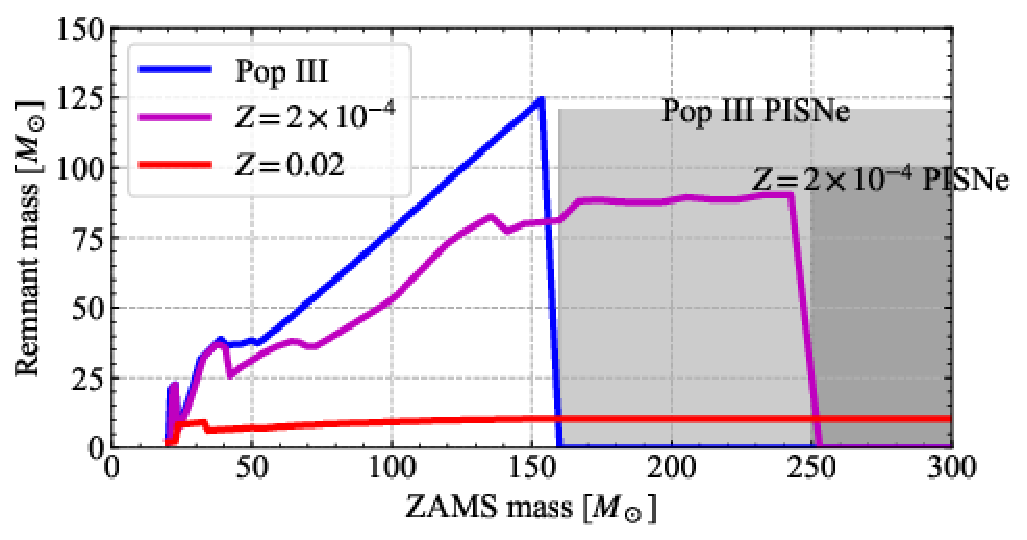}
\caption{\label{fig:zamsremnant3sg} Relation between ZAMS and remnant
  masses for $Z=\zsun$, $Z=10^{-2}\zsun$ and Pop III stars in the
  L-$\tsig$ set. Dark-gray and light-gray regions indicate that in
  which PISN is effective for Pop III and $Z=10^{-2}\zsun$ stars.
  $Z=\zsun$ stars do not experience PISN because of their small helium
  cores just before core collapse.}
\end{figure}

For the L-$\tsig$ set, we adopt the L model for the single star
evolution model, and the Fryer's rapid model for the supernova model,
which is the same as the L set. The L-$\tsig$ set has a different PI
mass loss from the L set. In the PI mass loss model, no PPISNe happen,
and PISNe are generated from stars with helium cores of $90$ -- $180
\msun$. PISN progenitor masses are shifted upward because of reducing
the $\cago$ reaction rate by $\tsig$ or 3 times. This is based on the
results of \cite{2020ApJ...902L..36F} and \cite{2021MNRAS.501.4514C},
and similar to the PI mass loss model adopted by
\cite{2020ApJ...905L..15B}.

Figure \ref{fig:zamsremnant3sg} shows the relation between ZAMS and
remnant masses for the L-$\tsig$ set. Pop III and $Z=10^{-2}\zsun$
stars cause PISNe for ZAMS masses of $\gtrsim 150 \msun$ and $\gtrsim
250 \msun$, respectively. $Z=10^{-2}\zsun$ stars need to have larger
ZAMS masses than Pop III stars, because they largely lose their masses
through their stellar winds. $Z=\zsun$ stars can not generate
PISNe. They cannot keep their masses large enough to cause PISNe just
before their core collapses.

Pop III stars with ZAMS masses of $\sim 150 \msun$ leave behind $\sim
125 \msun$ BHs. This is because they have their hydrogen envelopes
until their core collapses. For Pop III stars, PISN progenitors can
have hydrogen envelopes. On the other hand, $Z=10^{-2}\zsun$ stars
with ZAMS masses of $\sim 250 \msun$ leave behind $\sim 90 \msun$
BHs. They completely lose their hydrogen envelopes until their core
collapses. $Z=10^{-2}\zsun$ PISN progenitors are naked helium
stars. Actually, this is also true for the fid set. Observational
features of PISNe depend on whether hydrogen envelopes are present or
not \citep{2011ApJ...734..102K}.

\begin{figure}[b]
\includegraphics[width=0.5\textwidth]{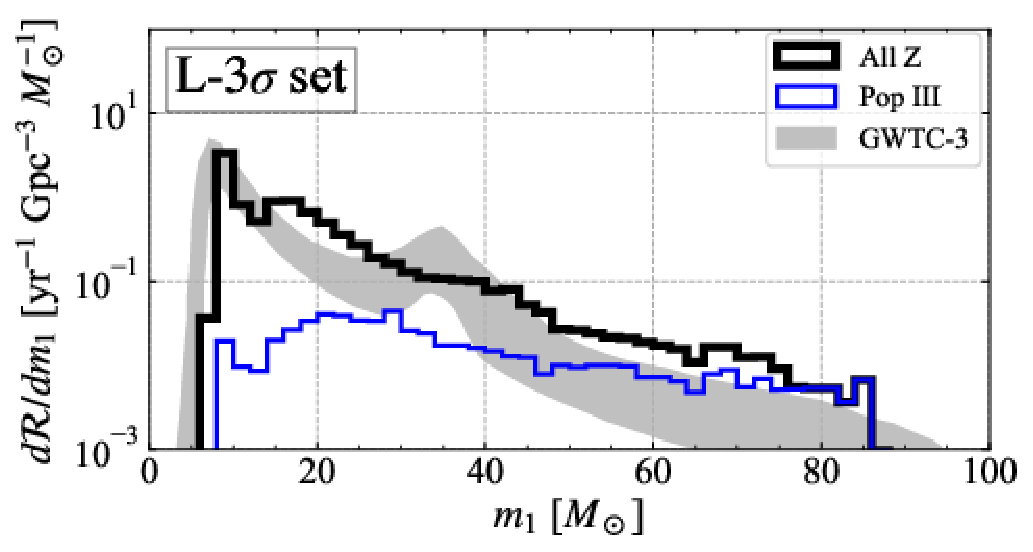}
\caption{\label{fig:primarymass3sg} Primary BH mass distribution of
  merging binary BHs at the present day for the L-$\tsig$ set. Black
  and blue curves indicate all the binary BHs and Pop III binary BHs,
  respectively. The gray shaded region indicates the corresponding
  primary BH mass distribution inferred by GWTC-3 within 90 \%
  credible bounds.}
\end{figure}

Figure \ref{fig:primarymass3sg} shows the primary BH mass distribution
of merging binary BHs at the present day for the L-$\tsig$ set. We can
see that the L-$\tsig$ set also produces merging binary BHs in the PI
mass gap. Such merging binary BHs can be yielded by both of Pop III
and non-Pop III binary stars. This reason is as follows. A Pop III
star with $\lesssim 150 \msun$ in a binary star loses its hydrogen
envelope through binary interaction and leaves behind a naked helium
star with $\lesssim 90 \msun$, because its maximum radius exceeds
$10^3 \rsun$ (see Figure \ref{fig:hrdpop3}). This naked helium star
can collapse to a BH with $\lesssim 90 \msun$. Note that PISNe only
happen for stars with helium core masses of $\gtrsim 90 \msun$.

\begin{figure}[b]
\includegraphics[width=0.5\textwidth]{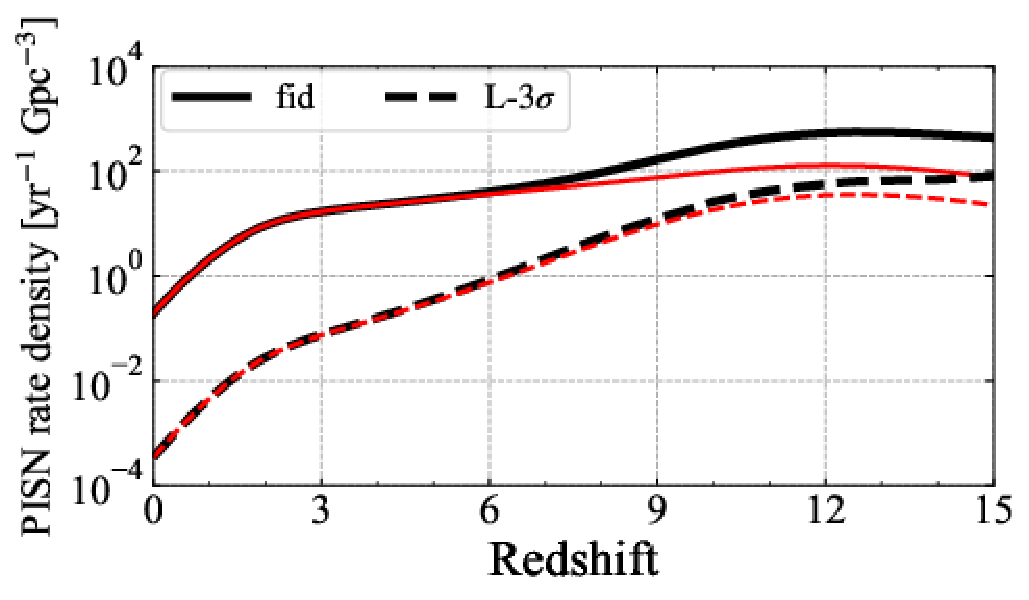}
\caption{\label{fig:pisnIntrinsicRateFromFile} Redshift evolution of
  intrinsic rate densities of PISNe for fid (solid) and L-$3\sigma$
  (dashed) sets. Black and red curves indicate the rate densities of
  all PISNe and PISNe with naked helium stars as progenitors.}
\end{figure}

We investigate different observational features between the fid and
L-$\tsig$ sets both of which can produce merging binary BHs in the PI
mass gap \citep{2023MNRAS.519L..32T}. We expect different PISN
features, because these sets have different PI mass loss models.

Figure \ref{fig:pisnIntrinsicRateFromFile} shows the redshift
evolution of PISN rate densities for the fid and L-$\tsig$ sets. Note
that we include not only binary stars but also single stars. For both
the sets, PISN rate densities increase with redshifts increasing up to
a redshift of $\sim 12$. For the higher-redshift universe, more
metal-poor stars are formed, and metal-poor stars cause PISNe more
easily because of weaker stellar winds. Up to a redshift of $\sim 6$,
PISN progenitors are dominated by naked helium stars. This is because
metal-rich stars (say $Z \gtrsim 10^{-2}\zsun$) lose their
hydrogen envelopes until their core collapses. Beyond a redshift of
$\sim 6$, PISN progenitors can have hydrogen envelopes, and a part of
PISN progenitors are Pop III stars.

For the L-$\tsig$ set, PISNe happen in spite of the fact that even Pop
III PISNe happen only when their ZAMS masses are larger than $150
\msun$ (see Figure \ref{fig:zamsremnant3sg}). Note that we set the
maximum ZAMS mass to $150 \msun$ as described in section
\ref{sec:InitialConditions}. For the L-$\tsig$ set, merged stars and
stars accreting mass from their companion stars cause PISNe. This is
partly because the PISN rate density is small. However, we make clear
that the PISN rate density increases only slightly even if we increase
the maximum ZAMS mass to $300 \msun$ \citep{2023MNRAS.519L..32T}. Star
cluster observations find at most $\sim 300 \msun$ stars
\citep{2013A&A...558A.134D, 2018A&A...618A..73S,
  2018Sci...359...69S}. $300 \msun$ ZAMS stars would be large enough
to be regarded as the maximum ZAMS mass.

We focus on the detectability of Euclid space telescope
\citep{2011arXiv1110.3193L}, because Euclid space telescope has a PISN
survey program \citep{2022A&A...666A.157M}. Euclid space telescope can
detect PISNe up to a redshift of $\sim 4$. The PISN rate density of
the fid set is larger than that of the L-$\tsig$ set by 2 orders of
magnitude. Actually, this difference has a large impact on the PISN
detectability by Euclid space telescope. For the fid and L-$\tsig$
sets, Euclid space telescope can detect two and zero PISNe,
respectively, for 5-year operation \citep{2023MNRAS.519L..32T}. Thus,
if Euclid space telescope will detect PISNe, the fid set may be
correct.

There was one concerning point for the above argument. This is because
there was no reference which describes observational features of PISNe
with the $\tsig$-smaller $\cago$ reaction rate. Recently, the amount
of Nickel-56 mass yielded from PISNe with various $\cago$ reaction
rates have been calculated \citep{2023arXiv230601682K}. Nickel-56
plays an important role in observational features of supernovae,
because its radioactive decay powers supernova luminosity. According
to \cite{2023arXiv230601682K}, Nickel-56 mass decreases with the
$\cago$ reaction rate decreasing if ZAMS mass is fixed. This means
that PISNe in the L-$\tsig$ set have not only a smaller rate density
but also smaller luminocities than those in the fid set. Thus, we can
confirm the above argument that no PISNe can be detected by Euclid
space telescope.

We have to make caveats on the above arguments. First, stellar wind
mass loss contains uncertainties, especially for massive stars causing
PISNe \citep{2021A&A...647A..13G, 2021MNRAS.504..146V}. This can
change ZAMS masses generating PISNe. Thus, the detection number of
PISNe for the fid set is changeable, although it is robust that the
detection number of PISNe for the fid set is much larger than for the
L-$\tsig$ set. Second, as described above, a part of PISNe are yielded
from merged stars and stars accreting masses from their companion
stars. In the BSEEMP model, merged stars and stars accreting masses
from their companion stars evolve in the same way as stars with the
same mass which do not experience mergers nor mass accretion. In fact,
merger and accretion history can alter stellar evolution. This is
because these effects cause efficient mixing of stellar interior,
providing fresh fuel \citep{1983Ap&SS..96...37H, 1984Ap&SS.104...83H,
  2021ApJ...923..277R}. These effects can enlarge helium core masses,
and finally affect PISN feasibility. Actually, our calculations
generate a large fraction of PISNe which do not happen without mergers
nor mass accretion. The fractions are $\sim 20$ and $\sim 50$ \% for
Pop III and $Z=10^{-2}\zsun$ stars, respectively. We may need to treat
merger and accretion history in more detail in order to obtain more
accurate PISN rate density.

\section{Summary}
\label{sec:Summary}

We have investigated features of merging binary BHs formed from
isolated binary stars over all metallicities from Pop III stars to the
solar-metal stars by means of rapid binary population synthesis
simulation. We have found that the redshift evolution of the binary BH
merger rate density in the fid set is consistent with the GWTC-3
result. The fid set can reproduce the primary BH mass and mass ratio
distributions inferred by the GWTC-3 result. Notably, Pop III binary
BHs dominantly contribute to merging binary BHs in the PI mass
gap. The binary BH population does not contain binary BHs with
negative effective spins. However, several studies have suggested that
binary BHs with negative effective spins in the GWTC-3 result might be
spurious. Thus, the absence of binary BHs with negative effective
spins in our result may not matter. Overall, our result successfully
reproduces the GWTC-3 result.

Our fid set can generate a large number of PISNe at the lower-redshift
universe, compared to the L-$\tsig$ in which binary BHs in the PI mass
gap can also merge within the Hubble time. If Euclid space telescope
will detect a few PISNe, we will conclude that the fid set is likely
to be correct, and that Pop III binary stars dominantly produce binary
BHs in the PI mass gap. Conversely, this means that we can use binary
BHs in the PI mass gap as a probe for Pop III stars. In any case,
ongoing and future observations will shed light on the origin(s) of
merging binary BHs and Pop III studies.

\section*{Conflict of interests}

On behalf of all authors, the corresponding author states that there
is no conflict of interest.

%\bibliographystyle{mnras}
%\bibliography{natbib} % if your bibtex file is called example.bib

\bsp	% typesetting comment
\label{lastpage}
\end{document}